\pdfoutput=1
\documentclass[useAMS,usenatbib]{mn2e}
\usepackage{cleveref}
\usepackage{natbib}
\usepackage{graphicx}
\usepackage{amsmath}
\usepackage{epstopdf}
\usepackage{subcaption}
\usepackage{float}

\newcommand{\msun}{\mbox{M}_\odot}

\newcommand{\au}{\,\mbox{AU}}

\newcommand{\Myr}{\,\mbox{Myr}}
\newcommand{\yr}{\,\mbox{yr}}
\newcommand{\cs}{\,c_\mathrm{s}}

\newcommand{\mH}{\,m_\mathrm{H}}

\def\lesssim{\mathrel{\hbox{\rlap{\hbox{\lower3pt\hbox{$\sim$}}}\hbox{\raise2pt\hbox{$<$}}}}}
\def\gtrsim{\mathrel{\hbox{\rlap{\hbox{\lower3pt\hbox{$\sim$}}}\hbox{\raise2pt\hbox{$>$}}}}}
\def\gtreq{\mathrel{\hbox{\rlap{\hbox{\lower3pt\hbox{$-$}}}\hbox{\raise2pt\hbox{$>$}}}}}
\def\lesssim{\mathrel{\hbox{\rlap{\hbox{\lower3pt\hbox{${\sim}$}}}\hbox{\raise2pt\hbox{$<$}}}}}

\title[Binary Star Formation and the Outflows from their discs]{Binary Star Formation and the Outflows from their Discs}
\author[Rajika L. Kuruwita, Christoph Federrath, Michael Ireland]{Rajika L. Kuruwita$^{1}$, Christoph Federrath$^{1}$, Michael Ireland$^{1}$\\ $^{1}$Research School of Astronomy and Astrophysics, Australian National University, Canberra, ACT 2611, Australia}
\begin{document}
\date{}

\pagerange{\pageref{firstpage}--\pageref{lastpage}} \pubyear{2015}

\maketitle

\label{firstpage}

\begin{abstract}
We carry out magnetohydrodynamical simulations with FLASH of the formation of a single, a tight binary ($a\sim$2.5$\au$) and a wide binary star ($a\sim$45$\au$). We study the outflows and jets from these systems to understand the contributions the circumstellar and circumbinary discs have on the efficiency and morphology of the outflow. In the single star and tight binary case we obtain a single pair of jets launched from the system, while in the wide binary case two pairs of jets are observed. This implies that in the tight binary case the contribution of the circumbinary disc on the outflow is greater than that in the wide binary case. We also find that the single star case is the most efficient at transporting mass, linear and angular momentum from the system, while the wide binary case is less efficient ($\sim$50$\%, \sim$33$\%, \sim$42$\%$ of the respective quantities in the single star case). The tight binary's efficiency falls between the other two cases ($\sim$71$\%, \sim$66$\%, \sim$87$\%$ of the respective quantities in the single star case). By studying the magnetic field structure we deduce that the outflows in the single star and tight binary star case are magnetocentrifugally driven, whereas in the wide binary star case the outflows are driven by a magnetic pressure gradient.
\end{abstract}


\section{Introduction}
\label{sec:introduction}

Stellar multiplicity is a critical output of star formation, with most solar mass or higher stars formed in multiple systems \citep{raghavan_survey_2010}. The fraction of all stars that are members of binary systems (multiplicity) depends on the mass of the system. For solar-type stars with effective temperature between $4800\,$K and $6550\,$K, the multiplicity of field stars is approximately $50$--$60$ per cent (\citealt{duquennoy_multiplicity_1991, kraus_mapping_2011}), while lower mass stars and dwarfs have a multiplicity of $30$ per cent down to $10$ per cent (\citealt{lada_stellar_2006, basri_survey_2006, ahmic_multiplicity_2007}). The prevalence of binary systems thus shows that a complete understanding of star and planet formation requires an understanding of binary formation. 

The current picture of star formation is that stars form preferentially in multiple star systems. The formation of binary star systems is believed to be triggered by the fragmentation of protostellar cores as they collapse. Each core produces around on average 2--3 stars due to fragmentation \citep{goodwin_limits_2005}. Models suggests fragmentation would occur via turbulent fragmentation (\citealt{klessen_fragmentation_1998, offner_formation_2010, offner_origin_2014}) or via gravitational instability in the protostellar disc (\citealt{takahashi_revised_2016, kratter_role_2010}). 
Individual star formation requires the loss of  $99$--$99.9\%$  of the initial angular momentum carried by the the molecular cores \citep{frank_jets_2014}, and an significant
fraction of this angular momentum can be stored in a multiple star system rather than lost.
Angular momentum can be carried out of the system via jets, outflows (there is no strict definition of the difference betweeen a jet and outflow, but a jet is generally considered to be a high velocity and strongly collimated outflow) or the ejection of a third companion. The formation of the binary star is then thought to broadly follow the single star birth.

During the formation of the binary, discs may form around individual components as circumprimary and circumsecondary discs, or around both stars as a circumbinary disc. The resonances produced by the binary system can greatly influence the evolution of the protoplanetary disc. \citet{artymowicz_dynamics_1994} showed in smoothed particle hydrodynamics (SPH) simulations that discs around each stellar component are truncated at the outer edge, and circumbinary discs are truncated along the inner edge. For a circular orbit, the circumstellar disc is truncated at $\sim a/2$, and the cirucmbinary disc is truncated at $\sim 2a$ where $a$ is the semi-major axis of the binary system. Greater eccentricities will lead to greater erosion of the discs, varying these values. The host binary may also affect the disc by exciting resonances that can encourage planet formation, by forcing gas and dust to fall into particular orbits \citep{bromley_planet_2015}. Binaries with $a < 40\au$ are half as likely to harbour circumprimary/circumsecondary discs than binaries with separations $40 - 400\au$ \citep{cieza_primordial_2009, duchene_planet_2010, kraus_role_2012}. This is also suggested by close binaries having less sub/millimetre flux due to absence of an inner disc \citep{jensen_connection_1994, jensen_connection_1996, andrews_circumstellar_2005}. 

Multiplicity may affect the disc lifetime, as the truncation of circumstellar discs by the companion may mean that disc material is accreted faster, on the order of $\sim$0.3~Myr \citep{williams_protoplanetary_2011}. Despite the assumption that circumbinary discs may have a shorter lifetime due to multiplicity, it has not been shown that cirucmbinary discs are short-lived compared to discs around single stars. There is also evidence for very long-lived circumbinary discs such as AK Sco \citep[$18\pm1\Myr$,][] {czekala_disk-based_2015}, HD 98800 B \citep[$10\pm5\Myr$,][]{furlan_hd_2007}, V4046 Sgr \citep[$12$--$23\Myr$,][]{rapson_combined_2015} and St 34 \citep[also known as HBC 425, $\sim 25 \Myr$,][]{hartmann_accretion_2005}.

The primary mechanisms for disc dispersal are: 1. accretion of material onto the stars, 2. photo-evaporation of the disc, 3. jets and outflows, which carry mass and momentum away from the system \citep{tomisaka_evolution_2000, tomisaka_collapse_2002} and are tied to the accretion history. Here we focus on how binarity affects the outflows and jets from young forming binary systems. The study of jet/outflows has so far focussed mostly on single stars. As most systems form binaries, it is important to quantify and understand the outflow properties and launching mechanisms in binary star systems.

To understand the outflows and evolution of binary stars we simulate the formation of a tight binary ($<10\au$) and a wide binary ($>10\au$), as well as a single star for reference. The setup and numerical methods of the simulations are described in \Cref{sec:method}. In \Cref{sec:results} and \ref{sec:discussion} we present the results from our simulations and discuss the impact binarity has on the outflows from the young systems. In \Cref{sec:conclusion} we summarise our results and present our conclusions.
	
\section{Method}
\label{sec:method}

\subsection{FLASH}
\label{ssec:flash}

The simulations are carried out with the magnetohydrodynamic (MHD) adaptive mesh refinement (AMR) code FLASH \citep{fryxell_flash:_2000, dubey_challenges_2008}. The standard set of ideal MHD equations are:  

\begin{equation}
\frac{\partial \rho}{\partial t} + \nabla \cdot (\rho \mathbf{v}) = 0,
\label{eqn:continuity}
\end{equation}

\begin{equation}
\rho \bigg( \frac{\partial}{\partial t} + \mathbf{v} \cdot \nabla \bigg) \mathbf{v} = \frac{(\mathbf{B} \cdot \nabla)\mathbf{B}}{4 \pi} - \nabla P_\mathrm{tot} + \rho \mathbf{g},
\label{eqn:momentum}
\end{equation}

\begin{equation}
\frac{\partial E}{\partial t} + \nabla \cdot \bigg( (E + P_{\mathrm{tot}}) \mathbf{v} - \frac{(\mathbf{B} \cdot \mathbf{v}) \mathbf{B}}{4 \pi} \bigg) = \rho \mathbf{v} \cdot \mathbf{g},
\label{eqn:energy}
\end{equation}

\begin{equation}
\frac{\partial \mathbf{B}}{\partial t} = \nabla \times ( \mathbf{v} \times \mathbf{B}),
\label{eqn:magneticfieldevolution}
\end{equation}

\begin{equation}
\nabla \cdot \mathbf{B} = 0.
\label{eqn:divB}
\end{equation}

\noindent Here $\rho, \mathbf{v}, \mathbf{B}, P_{\mathrm{tot}}, \mathbf{g}, E$ denote the gas density, velocity, magnetic field strength, total pressure (thermal plus magnetic), gravitational acceleration of the gas, and total energy density, respectively. Equations \ref{eqn:continuity}, \ref{eqn:momentum} and \ref{eqn:energy} describe the conservation of mass, momentum and energy, respectively, including the effects of magnetic fields and gravitational acceleration. The total pressure used is the sum of the magnetic pressure and the thermal pressure, where the magnetic pressure is given by:

\begin{equation}
P_{\mathbf{B}} = \frac{|\mathbf{B}|^2}{8\pi},
\label{eqn:magneticpressure}
\end{equation}

\noindent and the thermal pressure is defined by a piecewise polytropic equation of state (EOS),

\begin{equation}
P_\mathrm{th} = K\rho^\Gamma.
\label{eqn:eos}
\end{equation}

\noindent FLASH calculates the thermal pressure from this polytropic equation of state. Because of this, we technically do not use the internal energy from Equation \ref{eqn:energy}.

The $\Gamma$ used in our simulations is derived from \citet{masunaga_radiation_2000}:

\begin{equation}
\Gamma=\begin{cases}
1.0  \text{ for \,\,\,\,\,\,\,\,\,\,\,\, $\rho \leq \rho_1 \equiv 2.50 \times 10^{-16}$g$\,$cm$^{-3}$},\\
1.1  \text{ for $\rho_1 < \rho \leq \rho_2 \equiv 3.84 \times 10^{-13}$g$\,$cm$^{-3}$},\\
1.4  \text{ for $\rho_2 < \rho \leq \rho_3 \equiv 3.84 \times 10^{-8}$~g$\,$cm$^{-3}$},\\
1.1  \text{ for $\rho_3 < \rho \leq \rho_4 \equiv 3.84 \times 10^{-3}$~g$\,$cm$^{-3}$},\\
5/3 \text{ for \,\,\,\,\,\,\,\,\,\,\,\,$\rho > \rho_4$}.
\end{cases}
\label{eqn:gamma}
\end{equation}

These values approximate radiative transfer effects on the local cell scale in the gas including the initial isothermal contraction, adiabatic heating of the first core, the H$_2$ dissociation during the second collapse into the second core and the return to adiabatic heating. The polytropic constant used is $K = 4.0\times 10^8\,$cm$^2\,$s$^{-2}$. The value for the polytropic constant is derived from setting $K = \cs^2$, where $\cs$ is the sound speed. In the isothermal regime ($\Gamma = 1$) of our piecewise EOS, the sound speed is $\cs = 2\times 10^{4}\,$cm$\,$s$^{-1}$ for a temperature of $11\,$K for gas with mean molecular weight of $2.3\mH$ (where $\mH$ is the mass of a hydrogen atom).

The gravitational acceleration contains both the contribution from the gas and the sink particles, calculated using:

\begin{equation}
\mathbf{g} = -\nabla \Phi_\mathrm{gas} + \mathbf{g}_\mathrm{sinks},
\label{eqn:gravity}
\end{equation}

\noindent where $\Phi_\mathrm{gas}$ is the gravitational potential of the gas and $\mathbf{g}_\mathrm{sinks}$ is the gravitational acceleration from the sink particles (described in \Cref{ssec:sinkparticles}). FLASH integrates the ideal MHD equations such that the contribution from the gas and implemented sink particles is taken into account.

We use the HLL3R Riemann solver for ideal MHD \citep{waagan_robust_2011}. The gravitational interactions of the gas is calculated using a Poisson solver \citep{ricker_direct_2008}. The interactions between sink particles and the gas are computed using $N$-body integration.

\subsection{Sink Particles}
\label{ssec:sinkparticles}

In our simulations the formation of a protostar is signalled by the formation of a sink particle (\citealt{federrath_modeling_2010, federrath_implementing_2011, federrath_modeling_2014}). If a cell exceeds the density threshold, derived from the Jeans length, given by:

\begin{equation}
\rho_\mathrm{sink}= \frac{\pi\cs^2}{4Gr_\mathrm{sink}^2},
\label{eqn:densitythreshold}
\end{equation}

\noindent it may collapse (subject to additional checks; see below). The Jeans length must be resolved with at least four grid cells such that fragmentation is not artificial \citep{truelove_jeans_1997}. To prevent artificial fragmentation of a volume with radius $r_\mathrm{sink}=2.5 \Delta x$, where $\Delta x$ is the cell length on the highest AMR level, centred on the cell exceeding $\rho_\mathrm{sink}$ the volume must also meet the following criteria described by \cite{federrath_modeling_2010}. In order for a volume to create a sink particle the gas in the volume needs to:
\begin{enumerate}
	\item be on the highest level of grid refinement,
	\item not be within $r_\mathrm{sink}$ of an existing sink particle, 
	\item be converging from all directions ($v_\mathrm{r} < 0$),
	\item have a central gravitational potential minimum, 
	\item be bound ($|E_\mathrm{grav}| > E_\mathrm{th} + E_\mathrm{kin} + E_\mathrm{mag}$), and
	\item be Jeans-unstable.
\end{enumerate}
If a cell exceeds the density threshold while within $r_\mathrm{sink}$ of another sink particle, and bound and collapsing towards the sink, it will be accreted onto the sink particle. The accretion is carried out such that mass, momentum and angular momentum are conserved \citep{federrath_modeling_2014}. 

A second-order leapfrog integrator is used to update the particle positions using a velocity and acceleration based time step. A sub-cycling method is implemented to prevent artificial precession of the sink particles \citep{federrath_modeling_2010}.

\subsection{Simulation Setup}
\label{ssec:simulationsetup}

Using FLASH and the sink particle implementation described above, we simulate the formation of binary star systems and a single star for comparison of outflow quantities and structure of gas around the protostars. Because FLASH uses AMR to achieve higher resolution in regions with higher density, the effective number of cells (resolution) in each spatial dimension is defined as $2^L$ where $L$ is the level of refinement.

The size of the computational domain is $1.2\times 10^{17}\,$cm ($\sim$8000$\au$) along each side of the 3D computational domain. The minimum effective resolution in the simulation is $2^6$, thus one side has a minimum of 64 cells, with the cells being $\sim$125$\au$ across. The highest level of refinement in these simulations is $L=12$, making the highest resolution cells $\sim$1.95$\au$ across. At this resolution the accretion radius of sink particles is $r_\mathrm{sink}\sim$4.8$\au$. A convergence test has been carried out and the results are presented in Appendix A.

Our simulations begin with a large spherical cloud of mass $1\,\msun$, and radius $\sim$3300$\au$. The cloud is given solid body rotation with angular momentum of $1.85\times10^{51}\,$g$\,$cm$^2\,$s$^{-1}$. With this angular momentum, the product of the angular frequency and the freefall time of the cloud is $\Omega\times\,t_\mathrm{ff}=0.2$ (see \cite{banerjee_outflows_2006} and \cite{machida_high-_2008}). A magnetic field of $100\,\mu$G is also threaded through the cloud in the \emph{z}-direction. This gives a mass-to-flux fatio of $(M/\Phi)/(M/\Phi)_{\mathrm{crit}}=5.2$ where the critical mass-to-flux ratio is $487\,$g$\,$cm$^{-2}\,$G$^{-1}$ as defined in \citet{mouschovias_note_1976}. 

In the single star case (hereafter known as \emph{Single Star}) the cloud is given a uniform density of $\rho_0=3.82\times10^{-18}\,$g$\,$cm$^{-3}$. In the binary star cases a density perturbation is imposed on the cloud. This is to seed the formation of a binary star system. The density in this case is determined by:

\begin{equation}
\rho = \rho_0 [1 + \alpha_\mathrm{p}\mathrm{cos}(2\phi)],
\label{eqn:densityperturbation}
\end{equation}

\noindent where $\phi$ is the azimuthal angle around the $z$-axis and $\alpha_\mathrm{p}$ is the amplitude of the perturbation. For \emph{Single Star} $\alpha_\mathrm{p}$ is simply 0. For the tight binary simulation (hereafter known as \emph{Tight Binary}) $\alpha_\mathrm{p} = 0.25$ and for the wide binary simulations (hereafter known as \emph{Wide Binary}) $\alpha_\mathrm{p} = 0.50$. The cloud core rotates around the $z$-axis.

In the region outside the spherical cloud, there is gas density $\rho_0/100$ with a given internal energy such that the cloud and surrounding material is in pressure equilibrium. At the boundaries of our computational domain we use inflow/outflow boundary conditions.

We evolve three simulation setups described in \Cref{tab:simulation_summary}. The results of these three cases are presented in \Cref{sec:results}.

\begin{table}
	\centering
	\begin{tabular}{lcr}
		\hline
		Simulation name & $\alpha_\mathrm{p}$  & $a$ (AU)\\
		\hline
		\emph{Single Star} & 0.0 & - \\ 
 		\emph{Tight Binary} & 0.25 & $\sim2.5$ \\
 		\emph{Wide Binary} & 0.50 & $\sim45$ \\ 
		\hline
	\end{tabular}
	\caption{The middle column gives the density perturbation amplitude ($\alpha_\mathrm{p}$) as described by Eq. \ref{eqn:densityperturbation}. The right column gives the semi-major axis ($a$) of the resulting binary system formed in astronomical units. The semi-major axis is calculated from the last periastron and apastron passed shown in \Cref{fig:separation}}
	\label{tab:simulation_summary}
\end{table}

\section{Results}
\label{sec:results}

\begin{figure}
   \centerline{\includegraphics[width=1.0\linewidth]{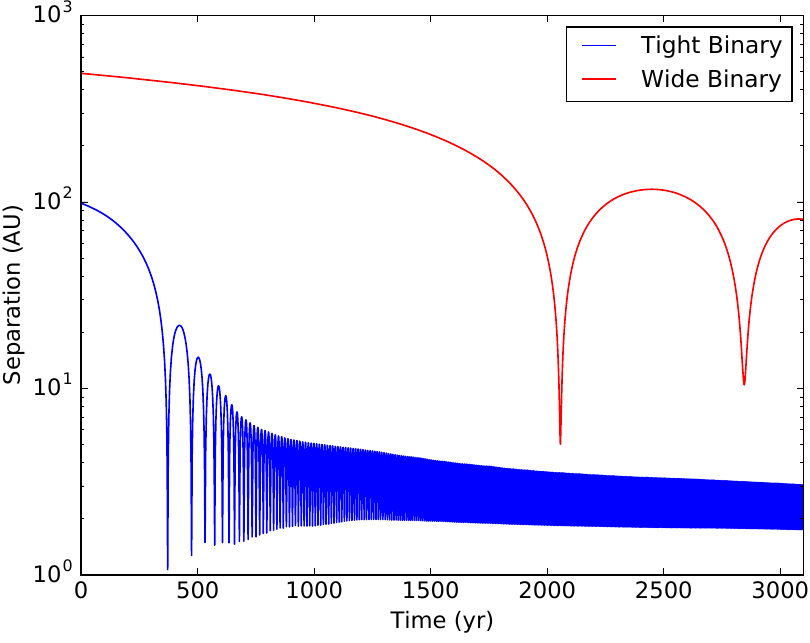}}
	\caption{Time evolution (since protostar formation) of the separation of the binaries, for the \emph{Tight Binary} (blue) and \emph{Wide Binary} (red) simulations.}
	\label{fig:separation}
\end{figure}

First, the morphology of the jets and outflows for each scenario (\emph{Single Star}, \emph{Tight Binary}, \emph{Wide Binary}) is discussed. Second, the quantitative analysis of the outflows is described and discussed. Lastly the identification of discs and jet launching are explored.

\subsection{Time evolution of the systems}
\label{ssec:evolution}

\begin{figure*}
\centerline{\includegraphics[width=1.0\linewidth]{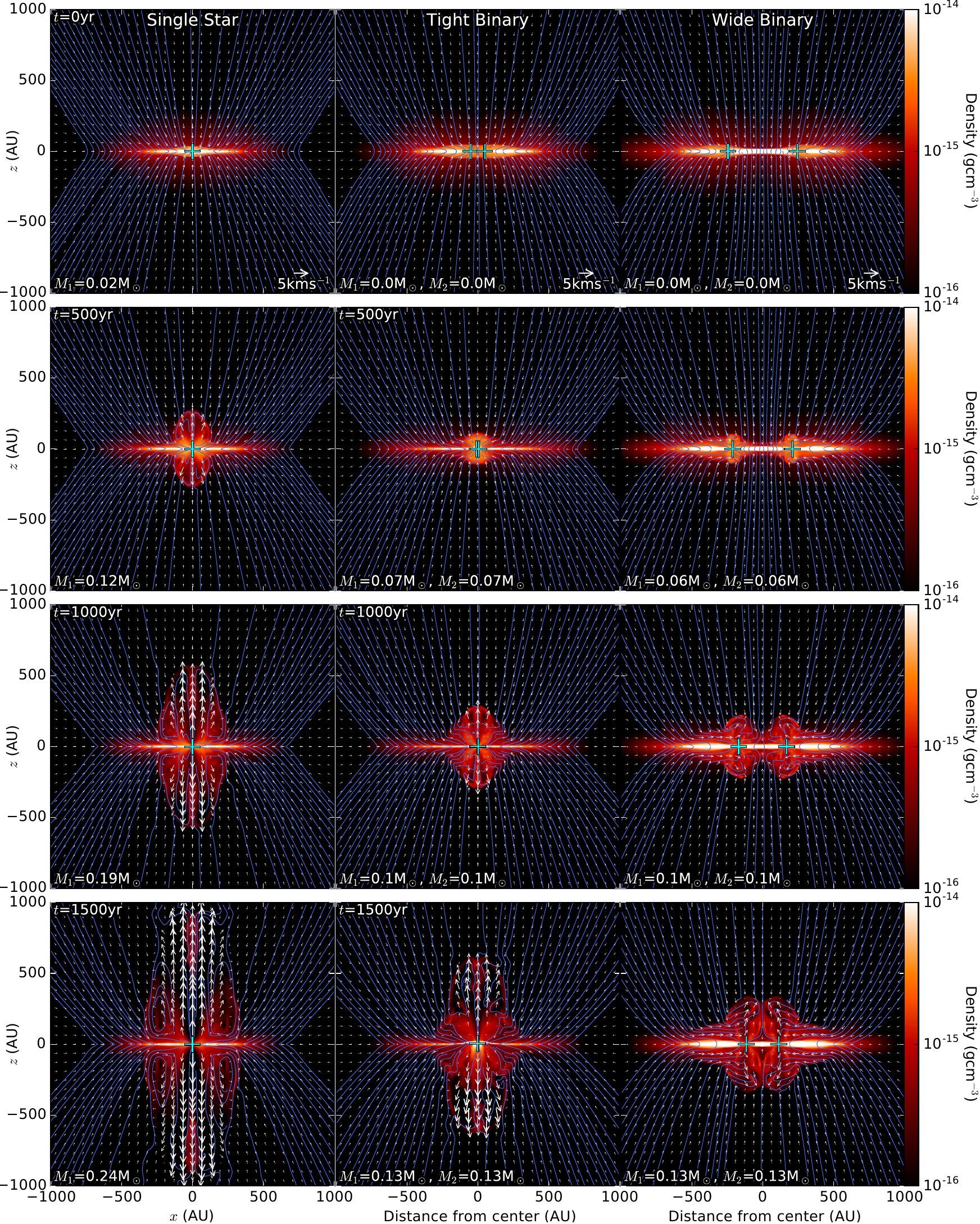}}
\caption{Slices along the $xz$-plane for \emph{Single Star} (left column) and along the separation axis such that the slice captures the two sink particles for \emph{Tight Binary} (middle column) and \emph{Wide Binary} (right column). Each row progresses at 500 year intervals since protostar formation. The thin lines show the magnetic field, and the arrows indicate the velocity field. Crosses show the position of the sink particles. The mass acceted by the sink particles in the simulations is indicated on the bottom left of each panel.}
\label{fig:side_on_slices}
\end{figure*}

As the simulations progress, the spherical cloud collapses and sink particles are created in collapsing regions (c.f. \Cref{ssec:sinkparticles}). In \emph{Single Star} a single sink particle forms at the centre of the computational domain. With the binary cases two sink particles form, which fall towards the centre of the computational domain. The initial separation that the sink particles have is dependent upon the strength of the density perturbation, with stronger perturbations creating wider initial separations. This is because sink particles form earlier and at wider separations.

The evolution of the sink particle separation in the \emph{Tight Binary} and \emph{Wide Binary} are presented in \Cref{fig:separation}. In \emph{Tight Binary} we see the particles form with a separation of $100\au$ and fall into a steady orbit after a few dozen orbits with semi-major axis $\sim$2.5$\au$. \emph{Wide Binary} forms particles with initial separation of $500\au$. This system does not fall into a steady orbit in the duration of our simulations.

\subsection{Morphology of the outflows}
\label{ssec:morphology}

Gas density slices through the $xz$-plane for \emph{Single Star} and along the separation axis for the binary star cases are shown in \Cref{fig:side_on_slices} at $500\yr$ intervals since the formation of sink particles. The left, middle and right column show slices for the \emph{Single Star}, \emph{Tight Binary} and \emph{Wide Binary} simulations, respectively. \Cref{fig:side_on_slices} qualitatively shows the outflow morphology in the three cases. \emph{Single Star} shows a strong jet breaking out of the disc, like in the simulations of \citet{federrath_modeling_2014}. \emph{Tight Binary} also produces a single jet which breaks out of the outflowing gas, but the launching time is delayed compared to that of \emph{Single Star}. The velocity of the \emph{Tight Binary} jet is also slower than that of \emph{Single Star}. \emph{Wide Binary} produces two individual jets, with launching time scales similar to \emph{Tight Binary}. These jets have even lower outflow velocities with respect to \emph{Single Star} and \emph{Tight Binary}. The outflow velocities for \emph{Single Star} and \emph{Tight Binary} are primarily perpendicular to the disc, whereas the outflow velocities of \emph{Wide Binary} bend towards the central axis of rotation of the binary system.

The jets in \emph{Tight Binary} are weaker relative to \emph{Single Star} likely because the inner edge of the circumbinary disc is truncated. The fastest components of outflows from centrifugally driven winds is launched from the innermost radii and the launching velocity of the outflows reduces with increasing radius \citep{blandford_hydromagnetic_1982}. However in circumbinary discs, this inner edge is truncated \citep{artymowicz_dynamics_1994}, reducing the velocity of outflows. \emph{Wide Binary} has the lowest outflow velocities from the two jets. This may be due to a different launching mechanism, which is discussed in \Cref{sec:launching_mechanisms}.

The outflow velocity of the jets in \emph{Wide Binary} may be skewed due to infalling material imparting momentum on the outflows pushing them inwards. We will also see in \Cref{sec:launching_mechanisms} that the magnetic field structure is strongly asymmetric, which may contribute to outflow velocities that are not perpendicular to the disc.

\subsection{Time evolution of outflow quantities}
\label{ssec:outflows}

\begin{figure}
\centerline{\includegraphics[width=1.0\linewidth]{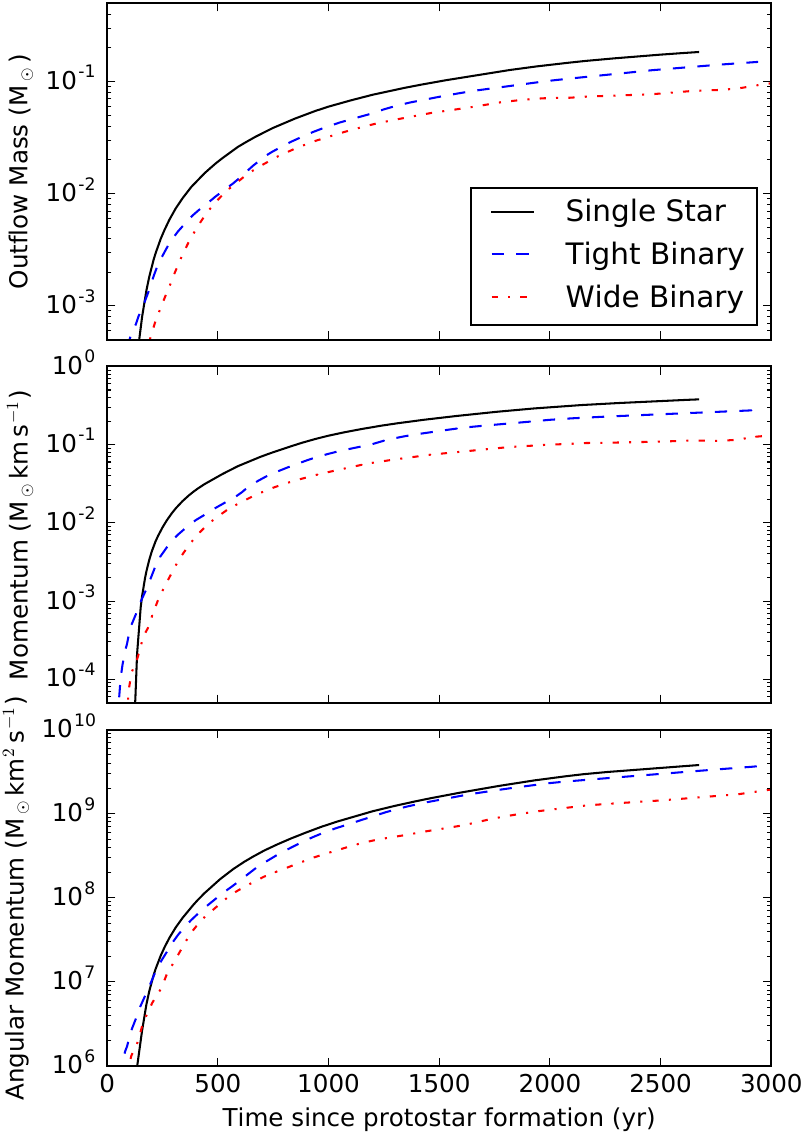}}
	\caption{Time evolution of the outflow quantities measured from outflowing material greater than two scale heights from the disc midplane (i.e. $|z| \geq 2z_\mathrm{H} = 50\au$). \emph{Top}: outflowing mass, defined as mass within cells with $v_z$ away from the disc. \emph{Middle}: linear momentum of the outflowing gas. \emph{Bottom}: angular momentum of the outflowing gas calculated around the centre of mass of the system.}
	\label{fig:outflow_quantities}
\end{figure}

\begin{figure}
\centerline{\includegraphics[width=1.0\linewidth]{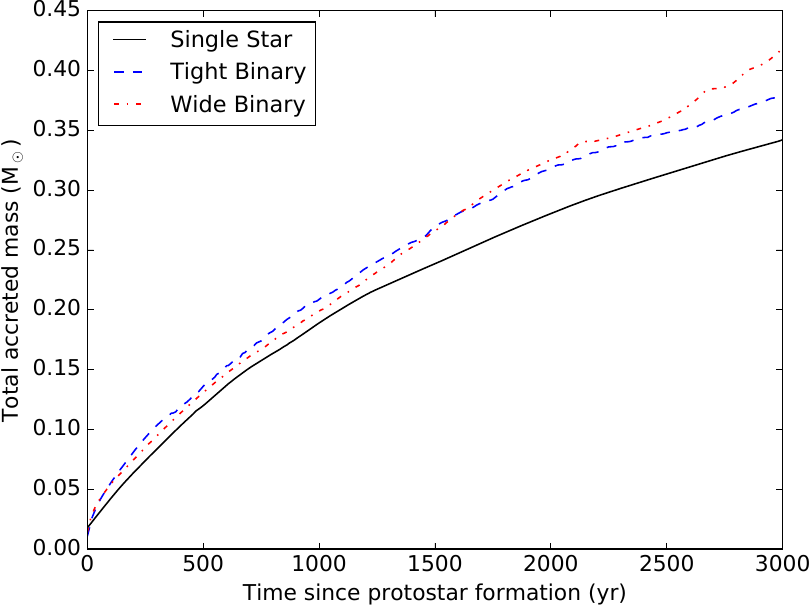}}
	\caption{The total mass accreted by the sink particles in \emph{Single Star} (solid line), \emph{Tight Binary} (dashed line) and \emph{Wide Binary} (dash dotted line). At all times the binary cases have accreted more mass compared to \emph{Sinlge Star}, which is likely the result of the respectively lower efficiency of the outflows in the binary cases compared to \emph{Single Star} (c.f.~\Cref{fig:outflow_quantities}).}
	\label{fig:total_mass}
\end{figure}

The outflows from a protostellar system are one of the mechanisms that carry away mass and momentum from the disc. Here we measure these outflows quantities in our three scenarios to determine how binarity may affect outflows from young systems.

The scale height of the discs in our simulations is $z_\mathrm{H} = 25\au$. Analysis of the outflows from the systems is carried out by measuring the outflowing mass where $|z| \geq 2z_\mathrm{H} = 50\au$. The momentum and angular momentum carried in the outflowing gas is also calculated. We decide to measure outflows from two scale heights from the disc to exclude most of the disc material. Within the measuring regions outflow mass is defined as any mass in cells with $v_z>0$ for $z>0$ and $v_z<0$ for $z<0$. From the outflow mass, the angular momentum and linear momentum of the outflows is calculated. The linear momentum is calculated from the magnitude of the velocity and the outflow mass. The angular momentum is calculated about the centre of mass of the systems. 

The outflow measurements of the three cases are shown in \Cref{fig:outflow_quantities}. In all cases \emph{Single Star} is the most efficient at transporting mass, momentum and angular momentum, while \emph{Wide Binary} is the least efficient in transporting those same quantities. In all cases there is already some outflow material by the time the sink particle forms.

To compare the overall outflow efficiency of these quantities we calculate the time averaged values of the quantity using:

\begin{equation}
<q> = \frac{\int_{0}^{T} q(t) dt}{\int_{0}^{T}dt},
\label{eqn:time_averaged}
\end{equation}

\noindent where $q$ is mass, momentum and angular momentum and $T=2500$~yr.

\emph{Single Star} is the most efficient at carrying mass in its outflows. The time averaged values of the outflow mass for \emph{Tight Binary} and \emph{Wide Binary} are $\sim$71$\%$ and $\sim$50$\%$ of \emph{Single Star}, respectively.

The outflow of \emph{Single Star} carries the greatest linear momentum. This is due to a combination of having the most massive outflows along with the highest outflow velocities. Similarly the outflows of \emph{Wide Binary} carry the least amount of linear momentum due to the same reasons. The time averaged values of the linear momentum for \emph{Tight Binary} and \emph{Wide Binary} are $\sim$66$\%$ and $\sim$33$\%$ of \emph{Single Star}, respectively.

\emph{Single Star} is the most efficient at transporting angular momentum via its outflows. It is expected that \emph{Single Star} is more efficient at transporting angular momentum from the disc compared to \emph{Wide Binary} because a binary star system has higher angular momentum than a single star with the same total mass. This is because greater angular momentum is necessary in order to maintain the binary orbit. The larger the binary separation, the greater the angular momentum necessary to maintain the orbit. As \emph{Tight Binary} has a very small separation, angular momentum of the system is comparable to \emph{Single Star}, hence the outflowing angular momentum of \emph{Tight Binary} converges towards the quantities measured in \emph{Single Star}. The time averaged values of the angular momentum for \emph{Tight Binary} and \emph{Wide Binary} are $\sim$87$\%$ and $\sim$42$\%$ of \emph{Single Star}, respectively.

\subsection{Evolution of the accreted mass}
\label{system_masses}

From the outflow analysis in \Cref{fig:outflow_quantities} we have determined that the binary cases are less efficient at transporting mass and momentum. Since the outflows are less efficiency in the binary cases compared to \emph{Single Star}, we expect that the binaries may have accreted more mass. \Cref{fig:total_mass} shows the total mass accreted by the sink particles in our three simulations. We see that the binary star cases have a greater star formation efficiency (fraction of accreted mass) than \emph{Single Star} at all times. \Cref{fig:total_mass} shows that $3000\yr$ after sink particle formation, the tight and wide binaries have respectively accreted $\sim$10$\%$ and $20\%$ more mass than the \emph{Single Star}. Moreover, the time evolution of the binary cases between $2500$ and $3000\yr$ after protostar formation indicates that the accretion rate increases compared to the \emph{Single Star}. However, we cannot predict the final mass of the stars, because it is impossible to run these simulations until the stars stop accreting mass, due to the limited amount of compute time currently available.

\subsection{Disc structure}
\label{ssec:discstructure}

\begin{figure*}
\centerline{\includegraphics[width=1.0\linewidth]{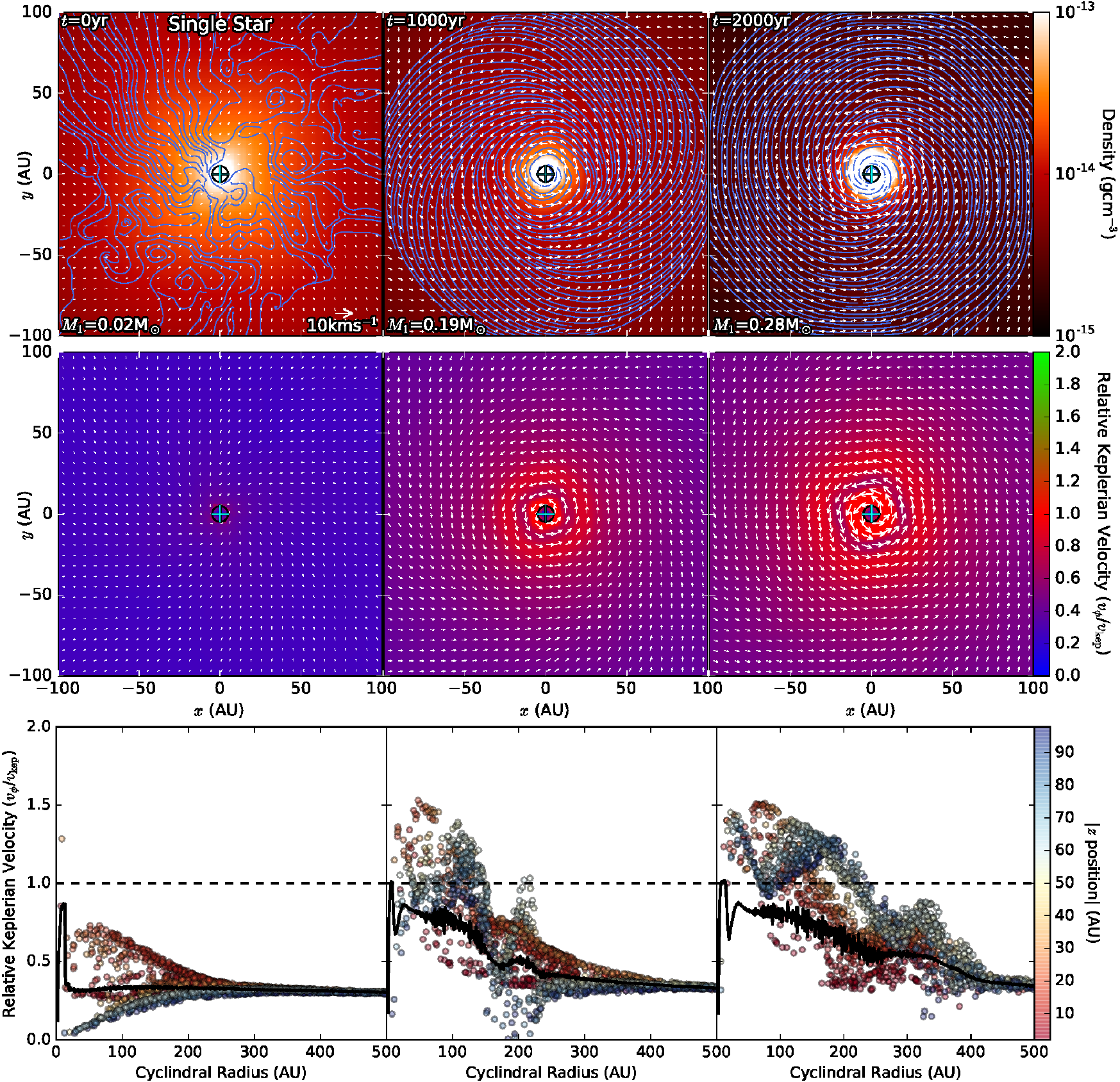}}
  \caption{\emph{Top}: slice plots of density for \emph{Single Star} in the disc midplane. The thin lines show the magnetic field, and the arrows indicate the velocity field. The centre of the cross marks the position of the sink particle, and the circle marks the accretion radius of the particle. 
Although the magnetic field in the $xy$ plane changes sign in the vicinity of the midplane, the streamline structure at 1000 and 2000 years has a near identical (spiral) structure just above and just below the mid plane.
  \emph{Middle}: slice plots of the relative Keplerian velocity defined in Equation \ref{eqn:relativeKeplerianvelocity}. \emph{Bottom}:  The solid black line is the mass-weighted radial profile of the relative Keplerian velocity for a cylindrical volume of radius $500\au$ and thickness $200\au$ centred on the centre of mass. The points are randomly selected cells in the volume at various distances from the mid-plane. Each column progresses at steps of $1000\yr$.}
  \label{fig:top_down_slices_single}
\end{figure*}

\begin{figure*}
\centerline{\includegraphics[width=1.0\linewidth]{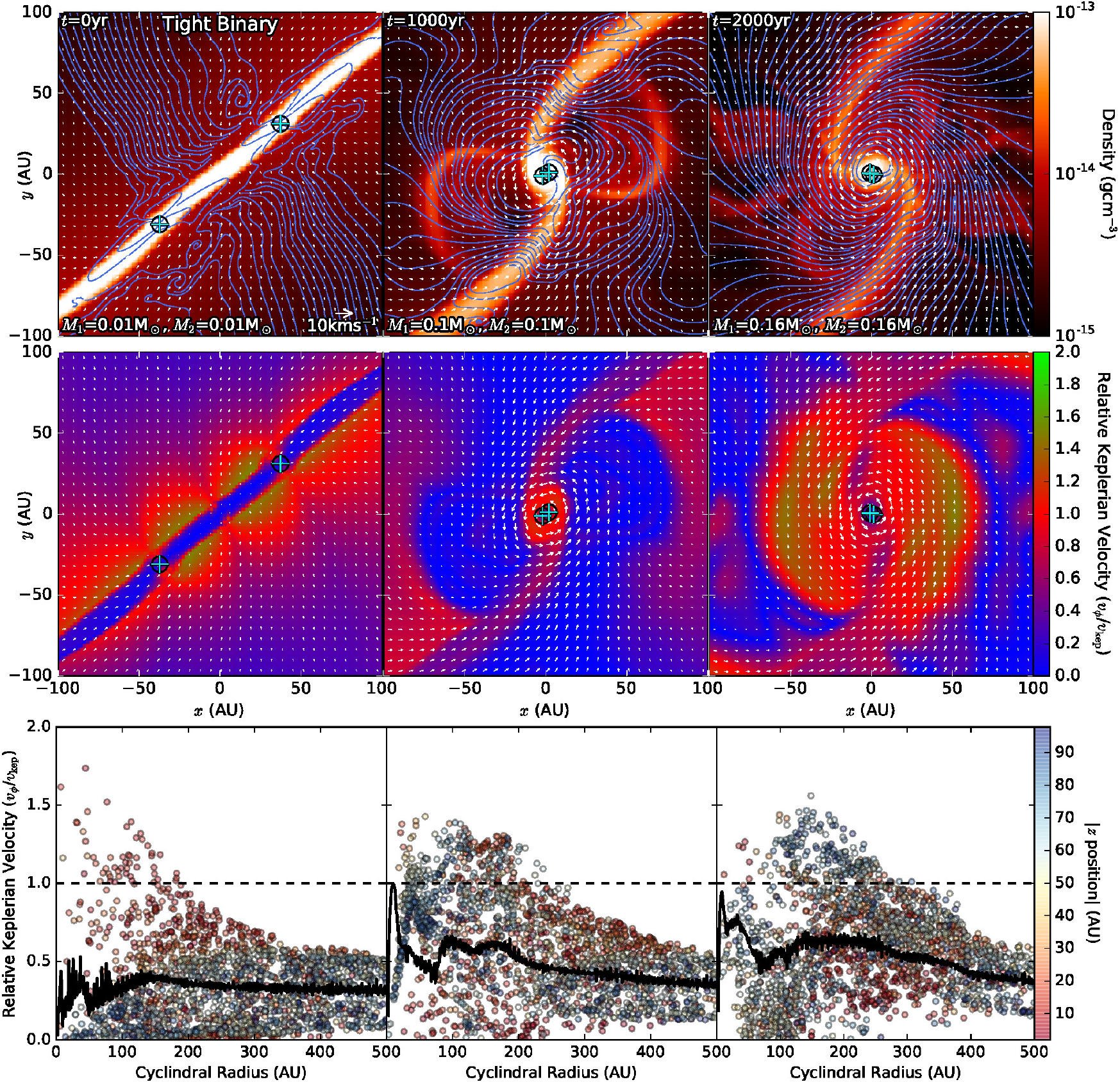}}
  \caption{Same as \Cref{fig:top_down_slices_single}, but for \emph{Tight Binary}.}
  \label{fig:top_down_slices_tight}
\end{figure*}

\begin{figure*}
\centerline{\includegraphics[width=1.0\linewidth]{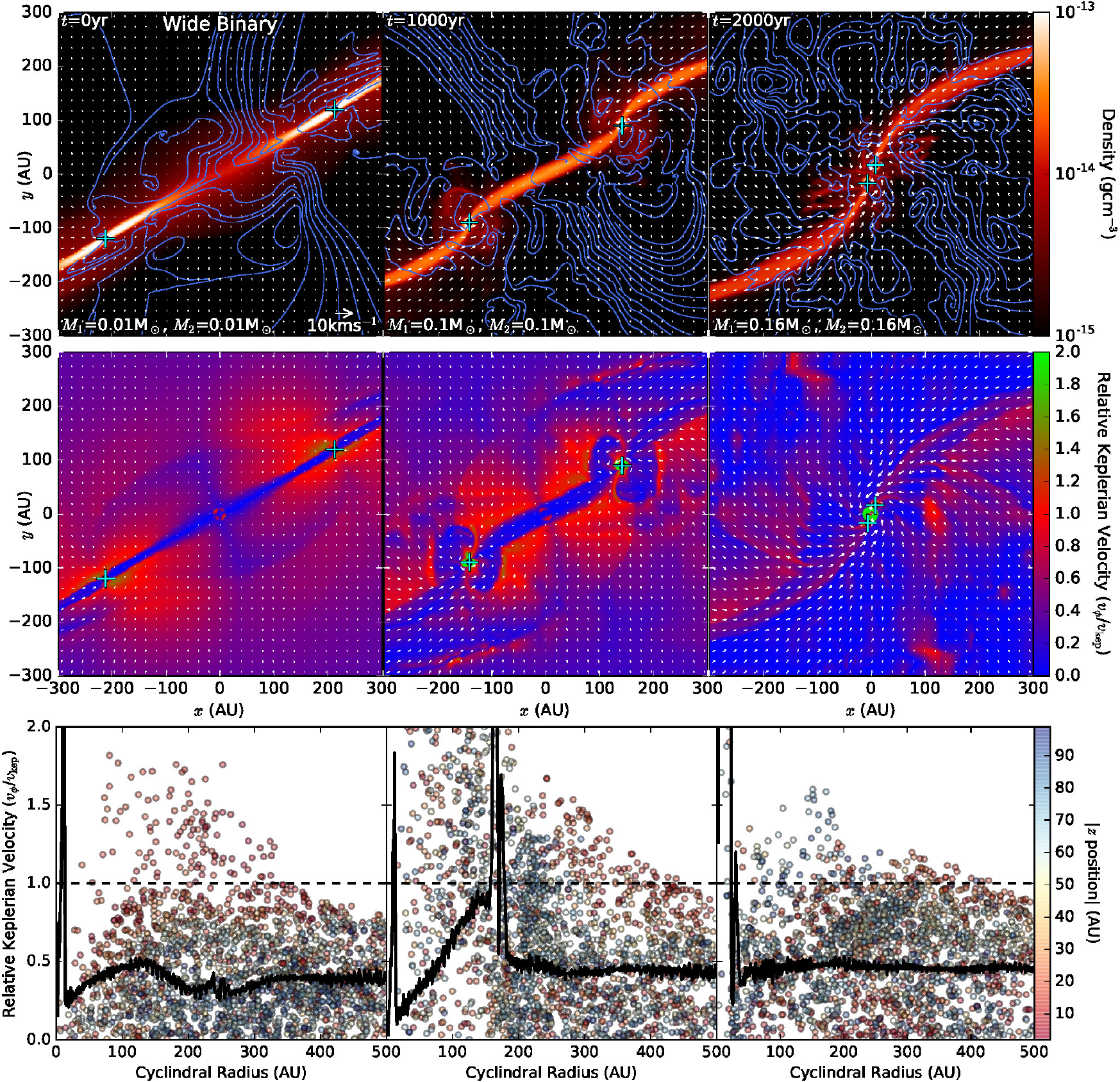}}
  \caption{Same as \Cref{fig:top_down_slices_single}, but for \emph{Wide Binary} zoomed-out to capture both sink particles.}
  \label{fig:top_down_slices_wide}
\end{figure*}

\begin{figure*}
\centerline{\includegraphics[width=1.0\linewidth]{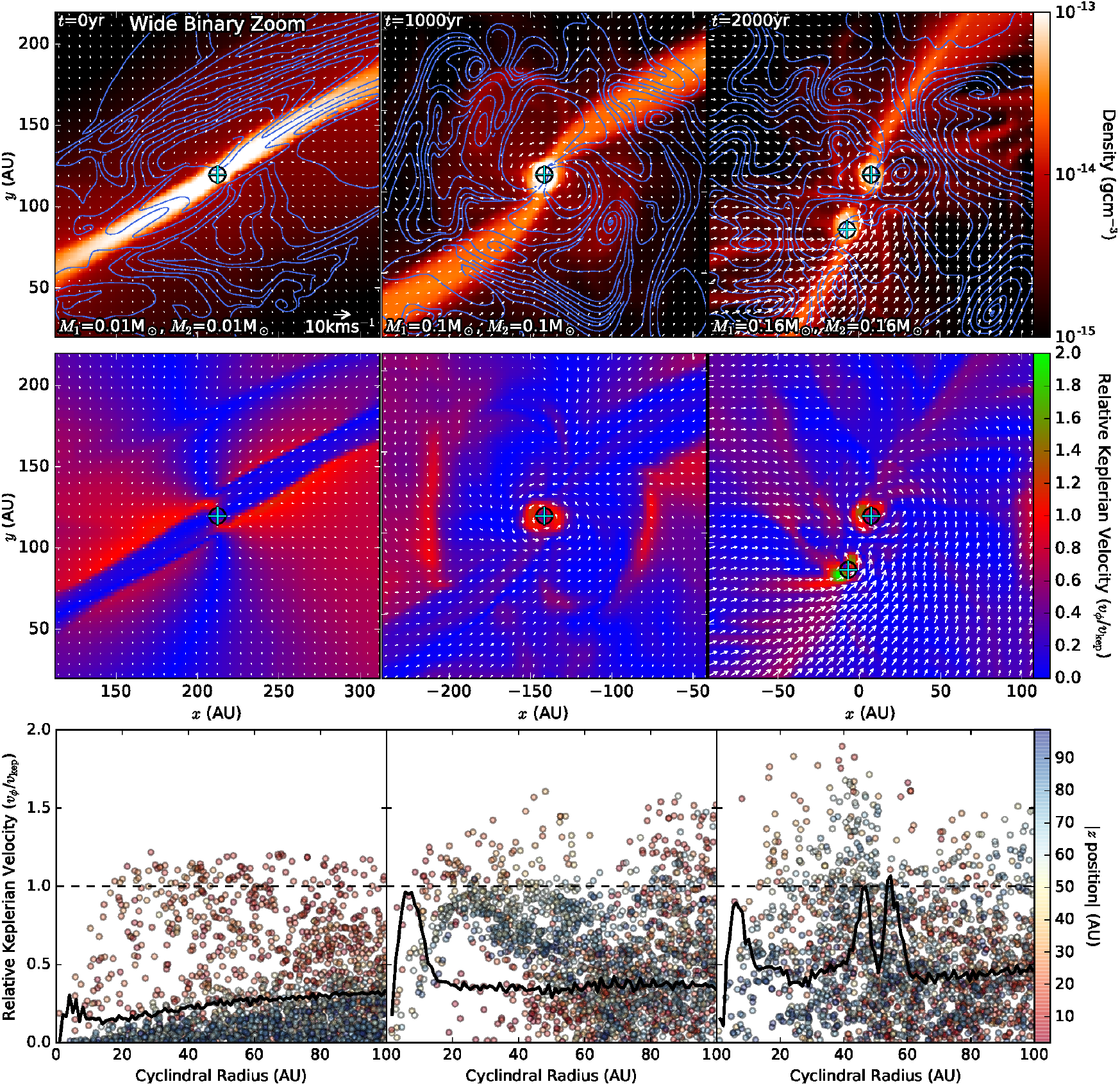}}
  \caption{Same as \Cref{fig:top_down_slices_single} for \emph{Wide Binary}, but centred on one sink particle.}
  \label{fig:top_down_slices_wide_zoom}
\end{figure*}

In Figures \ref{fig:top_down_slices_single}, \ref{fig:top_down_slices_tight} and \ref{fig:top_down_slices_wide} we present slices along the midplane of the density and relative Keplerian velocity for our three simulations at $1000\yr$ intervals. In the figures, the top panels show streamlines annotating the magnetic field. For clarification, the magnetic field exactly in the midplane is perpendicular to the slice. Here the annotated field is taken slightly off-centre from the midplane. We also show mass-weighted radial profiles of the relative Keplerian velocity. It should be noted that the data presented is very early in the evolution of the systems and we do not expect large Keplerian discs to be established so soon after protostar formation.

The formation of large Keplerian discs in ideal MHD simulations has been problematic due to the magnetic breaking catastrophe \citep{hennebelle_magnetic_2008}. Ideal MHD simulations of the collapse of magnetized cores of mass-to-flux ratios smaller than $\sim$5 (\citealt{hennebelle_magnetic_2008, mellon_magnetic_2008, joos_protostellar_2012}) do not produce Keplerian discs due to the magnetic field carrying away angular momentum very efficiently. This is despite observations indicating the presence of Keplerian discs around class 0 objects \citep{murillo_keplerian_2013} with radii of $\sim100\au$. Various solutions to this problem have proposed such as non-ideal MHD effects (\citet{duffin_early_2009}, see \Cref{ssec:non_iMHD_effects}) and turbulence (\citealt{seifried_turbulence-induced_2013, seifried_accretion_2015}). Our simulations have an initial mass-to-flux ratio of $\sim$4.8 which is near this threshold of $\sim$5 and we produce small Keplerian discs of a few AU. Other factors such as the initial angular momentum of the cloud core can contribute to the ability to produce Keplerian discs and the mass-to-flux threshold of $\sim$5 should not been taken as a rigorous criterion for establishing rotationally supported discs or not.

Because there are multiple centers about which the calculation of the Keplerian velocity can be taken (e.g. the centre of mass of the system and the individual sink particles), we find the Keplerian velocity about each possible centre. The final Keplerian velocity is taken to be the mean of the relative Keplerian velocity calculated about the centre of mass of the system and around the sink particles. The relative Keplerian velocity ($v_\mathrm{rel}$) is:

\begin{equation}
v_\mathrm{rel} = \frac{v_\phi}{v_\mathrm{kep}},
\label{eqn:relativeKeplerianvelocity}
\end{equation}

\noindent where $v_\phi$ is the tangential velocity to the radial direction and $v_\mathrm{kep}$ is the Keplerian velocity, given by: 

\begin{equation}
v_\mathrm{kep} = \sqrt{\frac{GM_\mathrm{enc}}{r}}
\label{eqn:Keplerianvelocity}
\end{equation}

\noindent where $M_\mathrm{enc}$ is the mass enclosed in a sphere with radius $r$ from the the chosen centre (either the centre of mass, or sink particle position).

For \emph{Single Star} the centre of mass is essentially the particle, therefore we do not expect these values to vary significantly when calculating the relative Keplerian velocity around the centre of mass and the sink particle. With the binary cases, the values may vary significantly close to the sink particles, but further out as the gravitational potential of the binary system begins to look like that of a single star, the values will converge. For each calculation of the Keplerian velocity the enclosed mass is the sum of the gas and sink particles within $r$.

The mass-weighted profiles (solid black line in bottom panel of \Cref{fig:top_down_slices_single}) are calculated over a cylinder of radius $500\au$ and thickness $200\au$ centered on the centre of mass of the systems. $1\au$ bins are used to produce these profile plots. Along with the radial profile, randomly selected cells are plotted to show any variation in relative Keplerian velocity perpendicular to the disc.

In \emph{Single Star} (\Cref{fig:top_down_slices_single}) the evolution of the disc density is axially symmetric. In the density slices we clearly see that the magnetic field is disorganised at time $0\yr$, but it coils around the sink particle as the system evolves. In the relative Keplerian velocity slices we see the material around the sink particle is Keplerian out to $\sim$50$\au$ (red colour) by $2000\yr$. The profile plot from time $0\yr$ shows that material above and below the midplane (blue scatter points) is severely sub-Keplerian compared to the material near the midplane (red scatter points) at the same radius. This indicates material is falling onto the circumstellar disc. After the jets and outflows are launched we see an inversion of the material in the mid-plane and the material above and below the disc. At time $2000\yr$ we see the material further away from the midplane has greater Keplerian velocity compared to that in the midplane. This is showing the infall of material along the midplane, while the material further away from the disc is leaving via outflows.

In \emph{Tight Binary} (\Cref{fig:top_down_slices_tight}) and \emph{Wide Binary} (\Cref{fig:top_down_slices_wide}) a dense bar forms due to our strong density perturbations similar to the bar fragmentation described by \citet{machida_first_2004} and \citet{machida_collapse_2005}. Along this dense bar two sink particles form to create the binary systems that we analyse here.

In \emph{Tight Binary} (\Cref{fig:top_down_slices_tight}) we see that the particles form and fall in along the dense stream. As the sink particles orbit each other, tendrils of material are flung away from the binary. At time $1000\yr$ the dense material nearest the sink particles is Keplerian in an elliptical disc extending to $\sim$15$\au$. The material in the streams is near Keplerian indicating that the mass is accreted onto the binary tangentially. The region outside the inner disc, barring the streams, is severely sub-Keplerian. At time $2000\yr$ the Keplerian to near super-Keplerain region has extended outwards to $\sim$50$\au$. This region does not correlate with any near-uniform density region, therefore it is difficult to determine whether a disc has been established.

In \emph{Tight Binary}, when the sink particles initially form, the magnetic field is predominantly perpendicular to the dense stream. When the sink particles have established a steady binary, we see that the magnetic field also coils around the binary system like that seen in \emph{Single Star}.

The profile plot of the \emph{Tight Binary} does not show a gradient between material in the midplane, and above and below the disc. At time $=0\yr$ the material above and below the disc (blue scatter points) is predominantly sub-Keplerian. This indicates infall of material onto the disc.

In \emph{Wide Binary} (\Cref{fig:top_down_slices_wide}) the sink particles form and fall in along a dense stream as in \emph{Tight Binary}. As the sink particles fall towards each other we see dense discs near the particles. These discs are confirmed in the zoomed in relative Keplerian velocity plot shown in \Cref{fig:top_down_slices_wide_zoom}, with the disc extending to $\sim$10$\au$. In \Cref{fig:top_down_slices_wide_zoom} the relative Keplerian velocity that is plotting in the second row is calculated just around the particle that the plots are centred on. These circumstellar discs are disrupted when the sink particles first pass by each other. Over the course of the \emph{Wide Binary} evolution the magnetic field remains mostly random and disorganised. The streams that the sink particles fall along are severely sub-Keplerian similar to \emph{Tight Binary}.

\section{Discussion}
\label{sec:discussion}

\begin{figure*}
\centerline{\includegraphics[width=1.0\linewidth]{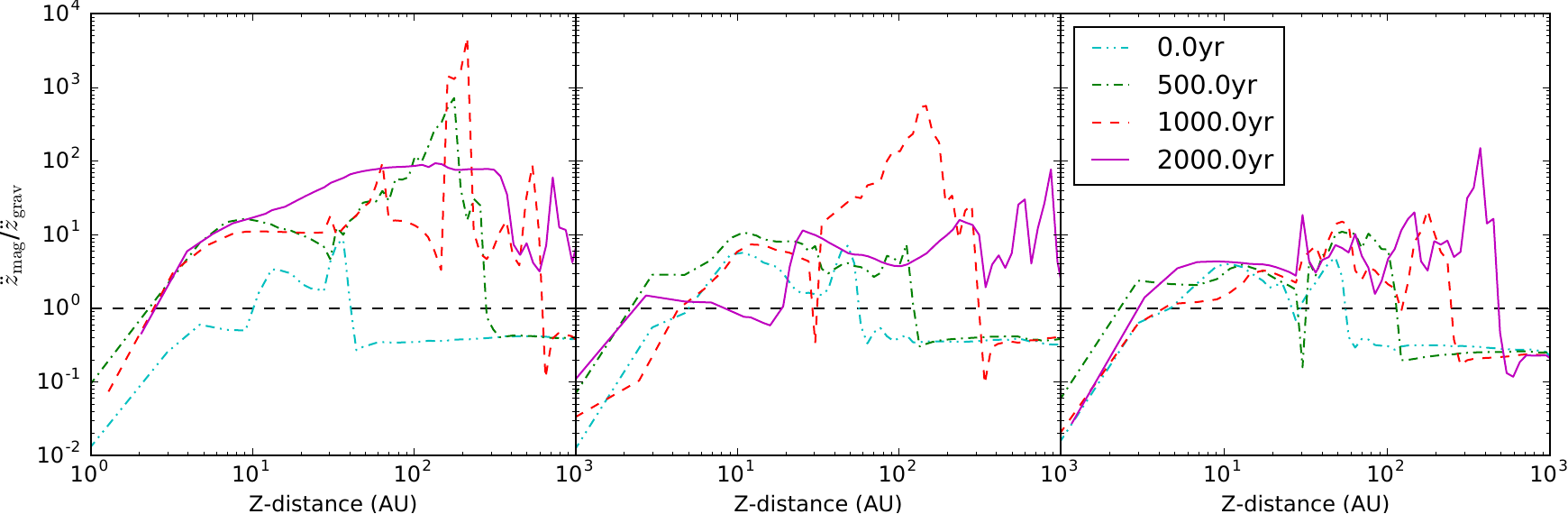}}
  \caption{Ratio of the acceleration due to magnetic pressure and gravitational potential for \emph{Single Star} (left), \emph{Tight Binary} (middle) and \emph{Wide Binary} (right) at times 0 years (dash-dash-dot cyan line), 500 years (dash-dot green line), 1000 years (dashed red line) and 2000 years (solid purple line) after sink particle formation.}
  \label{fig:acceleration_ratio}
\end{figure*}

\begin{figure*}
\centerline{\includegraphics[width=1.0\linewidth]{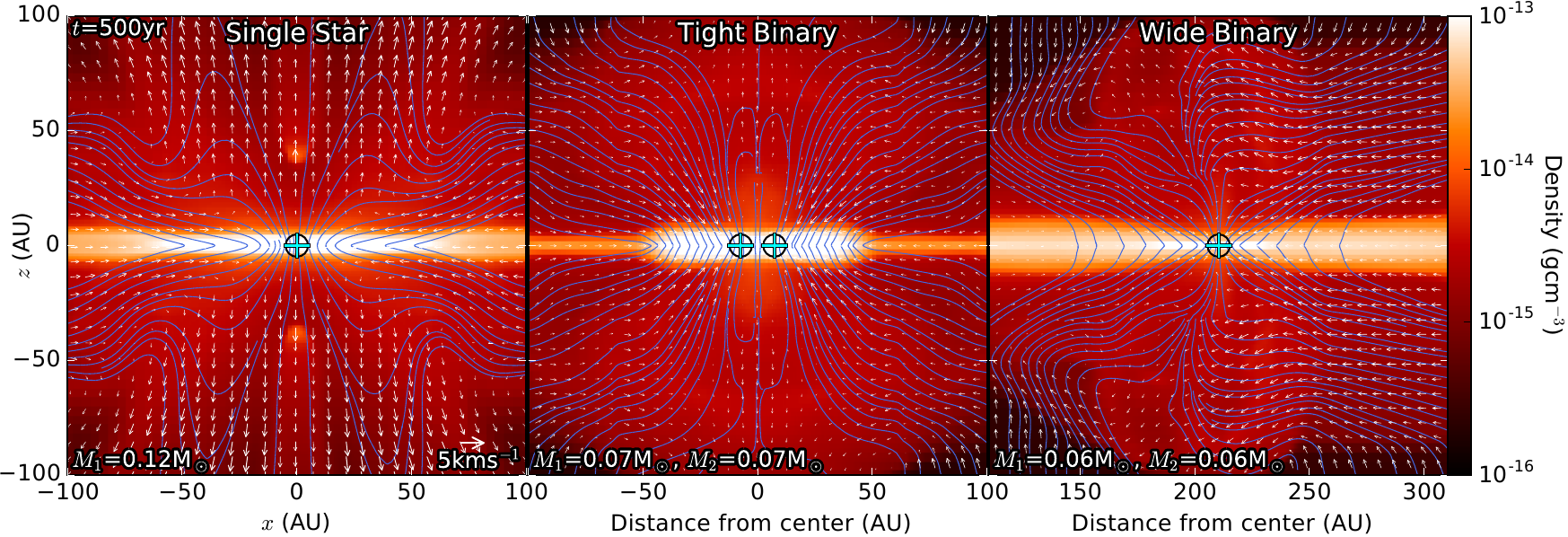}}
  \caption{Zoomed in plots of the slices of \Cref{fig:side_on_slices} at time $500\yr$ after sink particle formation.}
  \label{fig:side_on_zoom}
\end{figure*}

\begin{figure*}
\centerline{\includegraphics[width=1.0\linewidth]{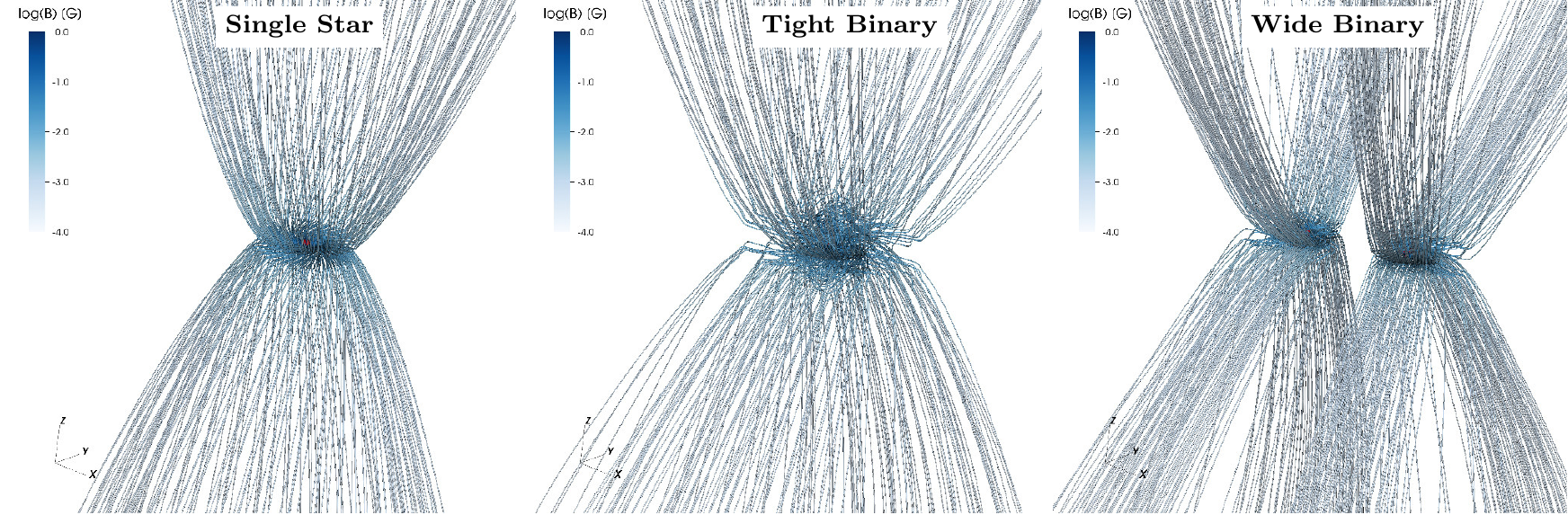}}
  \caption{3D streamline plots of the magnetic field structure around \emph{Single Star} (left), \emph{Tight Binary} (middle) and \emph{Wide Binary} (right). The colour bar shows the magnitude of the magnetic field.}
  \label{fig:3D_streamplots}
\end{figure*}

\subsection{Jet Launching Mechanisms}
\label{sec:launching_mechanisms}

Here we look at the outflow morphology and magnetic field structure to discuss how the outflows are launched in our three simulations.

From \emph{Single Star} we see one strong jet being produced. In \emph{Tight Binary} we also get one jet being produced. \citet{blandford_hydromagnetic_1982} describe magneto-centrifugally driven winds from protoplanetary discs which requires the winding up of magnetic fields around the protostar. In the top row of \Cref{fig:top_down_slices_single} and in the first panel of \Cref{fig:3D_streamplots} we see that the magnetic field lines are tightly wound about the sink particle at times $1000\yr$ and $2000\yr$. This suggests that this outflow is magnetocentrifgally driven. We also believe that the outflow from the \emph{Tight Binary} is driven in the same way. In the top row of \Cref{fig:top_down_slices_tight} we also see that the magnetic field does wind around the binary system at time $1000\yr$ and $2000\yr$, but not as tightly as \emph{Single Star}.

The magnetic field structure in the \emph{Wide Binary} is mostly disorganised compared to the other two cases. We do not see in \Cref{fig:top_down_slices_wide} large-scale winding up of the magnetic field that would indicate the presence of a circumbinary disc. However, \emph{Wide Binary} does have individual circumstellar discs which launch individual jets as seen in \Cref{fig:side_on_slices}. If the outflows from these individual discs are launched by the same mechanism that launches the outflows of the other two cases, we would expect greater launching velocities. But this scenario has the lowest outflow velocities.

\citet{lynden-bell_why_2003} describes outflows launched due to magnetic pressure gradients. Magnetocentrifugally driven outflows also require a magnetic pressure gradient which is provided by the tight winding of magnetic fields in the disc creating a denser magnetic field. Because we do not see the winding up of the magnetic fields around the individual sink particles in \emph{Wide Binary} in \Cref{fig:top_down_slices_wide_zoom}, we suggest that the outflows are launched from a general magnetic pressure gradient, weaker than that produced by the other two cases.

A stronger force due to the magnetic pressure gradient compared to the gravitational force is an indication of the strength of the outflow. To measure this for the three cases we calculate the ratio of the force due to magnetic pressure and the gravitational force in a cylindrical measuring volume of radius $25\au$ centered on a sink particle, extending to $1000\au$ above and below the disc. The direction of the acceleration is naturally in the opposite direction to the pressure gradient, which in our case is away from the disc. The magnetic pressure described by Equation \ref{eqn:magneticpressure} gives the magnetic pressure acceleration to be:

\begin{equation}
\ddot{z}_{\rm{mag}} = -\frac{1}{8\pi\rho}\frac{\rm{d}|\mathbf{B}|^2}{\rm{d}z}.
\label{eqn:magneticaccleration}
\end{equation}

\noindent where $\ddot{z}, \rho, |B|$ is acceleration in the $z$-direction, mass density and magnitude of the magnetic field. The gravitational acceleration is given by:

\begin{equation}
\ddot{z}_{\rm{grav}} = -\frac{GM_\mathrm{enc}}{z^2}.
\label{eqn:gravitationalaccleration}
\end{equation}

\noindent where $M_{\mathrm{enc}}$ is the enclosed mass. In our case the enclosed mass is calculated in spherical coordinates centered on the centre of mass.

The acceleration is calculated in 1000 bins along the $z$ axis. The force is then calculated using $F=m\ddot{z}$, where $m$ is the mass in the bin. The ratio of the magnetic force to the gravitational force for the three cases at time $0\yr$, $500\yr$, $1000\yr$ and $2000\yr$ is shown in \Cref{fig:acceleration_ratio}. We see that \emph{Single Star} has the largest force ratios near the disc. \emph{Tight Binary} and \emph{Wide Binary} have weaker force ratios. In each case we see spikes in the ratio that occur at greater distances from the disc at later times. These spikes in the force ratio are following the shock fronts of jets.

The ratio of the magnetic force to the gravitational force provides information on the velocity of the outflows but says little about the morphology of the outflows. In \Cref{fig:side_on_zoom} we show zoomed-in side-on plots of the three simulations at time $= 500\yr$ soon after outflows are beginning to be launched. The zoom in of \emph{Single Star} and \emph{Tight Binary} are centered on the centre of the simulation domain, while the zoom in of \emph{Wide Binary} is centered on one of the sink particles. We see in \emph{Single Star} and \emph{Tight Binary} that the magnetic field structure is symmetric about the rotation axis, while \emph{Wide Binary} is asymmetric with a magnetic field density gradient from larger radii to smaller radii. The symmetry of the magnetic field in \emph{Single Star} and \emph{Tight Binary} likely contribute to the symmetry of the outflow. From \Cref{fig:side_on_slices} we see that the velocity of the outflows are skewed towards the rotation axis of the system. This may be due to the magnetic field gradient seen across the sink particle pushing the outflows towards the rotation axis.

From these results we conclude that the structure of the magnetic fields contributes significantly to the symmetry and morphology of the outflows. The varying ratio of the magnetic force to the gravitational force between the three cases also support our hypothesis about the magnetocentrifual launching produces a stronger magnetic pressure gradient, leading to a greater acceleration of the outflows.

\subsection{Caveats}
\label{ssec:caveats}

\subsubsection{Numerical resolution}
\label{ssec:resolution}

In Appendix A we show the results of our resolution study. \citet{federrath_modeling_2014} find that fully converged values require a level of refinement of $L\gtrsim17$ for a computational domain the same size as that used in our simulations (giving the highest resolution cells a length of $\Delta x =0.06\,\au$). A level of refinement of $17$ is very computationally intensive, therefore we used a lower level of refinement. The primary goal of this paper is to relatively compare our three scenarios, therefore we are not concerned about being fully converged.

We use a level of refinement of $L =12$, which gives the smallest cell size to be $\sim1.95\au$. This causes some issues resolving structure near the sink particles. In \emph{Tight Binary} we only discuss outflows originating from the circumbinary disc. This is because our simulations could not fully resolve individual circumstellar discs around the sink particles. Following the results of \citet{artymowicz_dynamics_1994}, if \emph{Tight Binary} had individual circumstellar discs they would be no larger than $1\au$ in radius.

Using a higher level of refinement would allow us to resolve the structure nearest to the sink particles in the disc and would lead to resolving greater launching velocities of the outflow.

\subsubsection{Radiation effects}
\label{ssec:radiation_effects}

We do not explicitly calculate radiative transfer in our simulations, however the equation of state that we use approximates some of these radiative effects on the local cell scale as described in \Cref{ssec:flash}. 

Radiative feedback is most important when considering massive stars and their radiation driven winds as well as accretion heating. \citet{kuiper_protostellar_2016} find that radiative feedback is reduced when the protostar initially forms with low luminosity and when outflows clear bipolar cavities. However, \citet{stamatellos_episodic_2012} ran smoothed particle hydrodynamic (SPH) simulations of episodic accretion onto low-mass stars with radiative feedback and found that fragmentation of the disc was suppressed. However, provided there was sufficient time between accretion events ($\geq 1000$~yr) and a low quiescent accretion rate ($\sim$10$^{-7}$~M$_\odot$~yr$^{-1}$) the discs could fragment to produce companions. 

There have been plenty of previous simulations concerning binary star formation that have included radiative transfer and feedback (\citealt{offner_formation_2010, bate_stellar_2012, buntemeyer_radiation_2016, kuiper_protostellar_2016}). \citet{offner_formation_2010} found that the inclusion of radiative transfer in the flux-limited diffusion approximation in hydrodynamic simulations leads to binary stars forming predominantly from core fragmentation rather than disc instabilities. Our initial conditions are designed such that our binary star systems form from core fragmentation, in agreement with the results of \citet{offner_formation_2010}. \citet{bate_stellar_2012} ran radiation hydrodynamical simulations of star cluster formation and conclude the main physical processes involved in determining the properties of multiple stellar systems are gravity and gas dynamics.

There have already been works considering both radiative feedback and ideal MHD mostly concerning cluster formation (\citealt{offner_effects_2009, price_inefficient_2009, myers_fragmentation_2013, myers_star_2014, krumholz_what_2016}). These works found that radiation mainly contributed to the suppression of fragmentation of the cloud.  \citet{bate_stellar_2012} and \citet{krumholz_what_2016} found that radiation greatly hindered the formation of brown dwarfs of $\sim$0.01$~$M$_\odot$. This is because the surrounding material is likely to be accreted as thermal pressure in the vicinity of the protostar prevents further fragmentation. This could also contribute to why low-mass binary star systems are less common than high-mass binary star systems.

Overall the effect that radiative feedback may have on our simulations is that it would help suppress fragmentation. The effects should be investigated in future work.

\subsubsection{Non-ideal MHD effects}
\label{ssec:non_iMHD_effects}

Non-ideal MHD effects are most important where partial ionisation fractions would occur such as in protoplanetary discs. The non-ideal effects of Ohmic resistivity, the Hall effect and ambipolar diffusion are important at $\sim$1.5, $2-3$ and $\geq 3$ scale heights respectively (\citealt{wardle_magnetic_2007, salmeron_magnetorotational_2008, konigl_effects_2011}). At greater scale heights the surface layers of discs are expected to be ionized by stellar radiation in the FUV and the ideal MHD limit is a reasonable approximation \citep{perez-becker_surface_2011}. 

These non-ideal MHD effects on star formation have been studied in detail (\citealt{duffin_simulating_2008, duffin_early_2009, gressel_global_2015, wurster_can_2016, bai_hall_2017, wurster_impact_2017}). \citet{wurster_can_2016} conducted a comprehensive study of individual non-ideal MHD effects on the collapse of cloud cores with SPH. Ohmic resistivity had no major effect on the morphology of the outflows or size of the protoplanetary disc. The Hall effect allowed for a jet component to form, but suppressed the formation of a disc if the initial angular momentum vector ($\mathbf{L}_0$) and the magnetic field ($\mathbf{B}_0$) were aligned ($\mathbf{L}_0 \cdot \mathbf{B}_0 > 0 $). However, if $\mathbf{L}_0 \cdot \mathbf{B}_0 < 0 $ the formation of an outflow was suppressed in their simulations, but a large disc was established. Ambipolar diffusion created a less dense jet and disc component, and changed the shape of the jet's head compared to the ideal MHD case. In the case including all non-ideal MHD effects, for $\mathbf{L}_0 \cdot \mathbf{B}_0 > 0 $ (like that in our simulations) their simulations created a less dense jet and disc because ambipolar diffusion was the most dominant non-ideal MHD effect. This is likely because the regions where ambipolar diffusion is most important (large heights from the disc midplane) are the most resolved in these simulations. This is supported by \citet{gressel_global_2015} who also found that ambipolar diffusion contributed the most to the outflows in their disc simulations which included Ohmic resistivity and ambipolar diffusion. The simulations of \citet{gressel_global_2015} with only Ohmic resistivity produced results similar to ideal MHD.

There has already been some work looking at non-ideal effects specifically on binary star formation. \citet{machida_circumbinary_2009} looked at the influence of Ohmic resistivity on outflows from the circumbinary disc around a tight $\sim$5 - 10\,AU binary. They found that Ohmic resistivity weakens the magnetic field strength near the protobinary but accumulated in the circumbinary disc where they are launched. \citet{duffin_early_2009} and \citet{wurster_impact_2017} found that these effects do not have significant impact on the overall early evolution of the binary stars. \citet{wurster_impact_2017} found that when Ohmic resistivty, the Hall effect and ambipolar diffusion were all included in their simulations of binary star formation, more massive discs were formed and the binary forms on a wider orbit. They also found that non-ideal MHD effects were amplified by the binary interaction near periastron, but overall non-ideal effects have little influence on binary formation and the initial conditions played the dominant role.

Our work does not consider non-ideal MHD effects because we are mainly concerned with the comparison between the formation of a single star and binary stars. Given the results of previous studies on binary star formation and outflows produced from accretion discs, the relative differences between \emph{Single Star}, \emph{Tight Binary} and \emph{Wide Binary} determined here, are not expected to change significantly with the inclusion of non-ideal MHD effects. The relative differences between these cases are not expected to change significantly with the inclusion of non-ideal MHD effects.

\section{Summary and Conclusion}
\label{sec:conclusion}

We ran and analysed MHD simulations of the formation of a single star, a tight binary and wide binary star system. We quantified the accretion, the outflow mass, momentum and angular momentum as well as the morphology of the outflows in these three simulation cases. We find the following main results.

\textbf{Outflow Morphology}: Our simulations produce jets and outflows not only from our single star case (as shown in previous simulations), but also from both our binary star cases. \emph{Single Star} and \emph{Tight Binary} produce a single jet, whereas the \emph{Wide Binary} simulation produces two distinct outflows. The morphology of the \emph{Single Star} and \emph{Tight Binary} outflows look similar apart from the \emph{Tight Binary} having low outflow velocities (c.f.~\Cref{fig:side_on_slices}).

\textbf{Outflow efficiencies}: The \emph{Single Star} simulation produces outflows with the greatest efficiency in terms of transporting mass, linear momentum and angular momentum away from the sink particle and disc system. \emph{Tight Binary} produces outflows carrying angular momentum with efficiencies that converge towards that measured in \emph{Single Star}. \emph{Wide Binary} has the weakest outflow efficiencies in mass, linear and angular momentum (c.f.~\Cref{fig:outflow_quantities}). However, as a result of having weaker efficiencies the binary star cases have accreted more mass than \emph{Single Star} (c.f.~\Cref{fig:total_mass}). This implies that the star formation efficiency is greater in the binary star cases than in \emph{Single Star}.

\textbf{Launching mechanisms}: Based on the winding up of magnetic fields around the sink particles in \emph{Single Star} and \emph{Tight Binary} we suggest that the outflows are launched via a magnetocentrifugal mechanism. The magnetic field winding up in the circumbinary disc in \emph{Tight Binary} leads to the production of only one jet from this system. By contrast, the structure of the magnetic field in \emph{Wide Binary} is disorganised, not showing the strong evidence of winding up seen in the other two cases. As a result we suggest the outflows in \emph{Wide Binary} are launched by a general magnetic pressure gradient (c.f.~Figures~\mbox{\ref{fig:top_down_slices_single}--\ref{fig:top_down_slices_wide_zoom}}).

Overall, when looking at morphology of outflows a tight binary star system may look like a single protostar but the outflows are weaker and less efficient.

\section*{Acknowledgments}

We thank the anonymous referee for the constructive feedback and comments that helped to improve the paper. R.K.~would like to thank the Australian Government and the financial support provided by the Australian Postgraduate Award. C.F.~gratefully acknowledges funding provided by the Australian Research Council's Discovery Projects (grants~DP150104329 and~DP170100603). The simulations presented in this work used high performance computing resources provided by the Leibniz Rechenzentrum and the Gauss Centre for Supercomputing (grants~pr32lo, pr48pi and GCS Large-scale project~10391), the Partnership for Advanced Computing in Europe (PRACE grant pr89mu), the Australian National Computational Infrastructure (grant~ek9), and the Pawsey Supercomputing Centre with funding from the Australian Government and the Government of Western Australia, in the framework of the National Computational Merit Allocation Scheme and the ANU Allocation Scheme. 
The simulation software FLASH was in part developed by the DOE-supported Flash Center for Computational Science at the University of Chicago.

\bibliographystyle{mn2e}
\bibliography{Bibliography}

\appendix
\section{Convergence Tests}
\label{sec:appendix}

\begin{figure*}
\centerline{\includegraphics[width=1.0\linewidth]{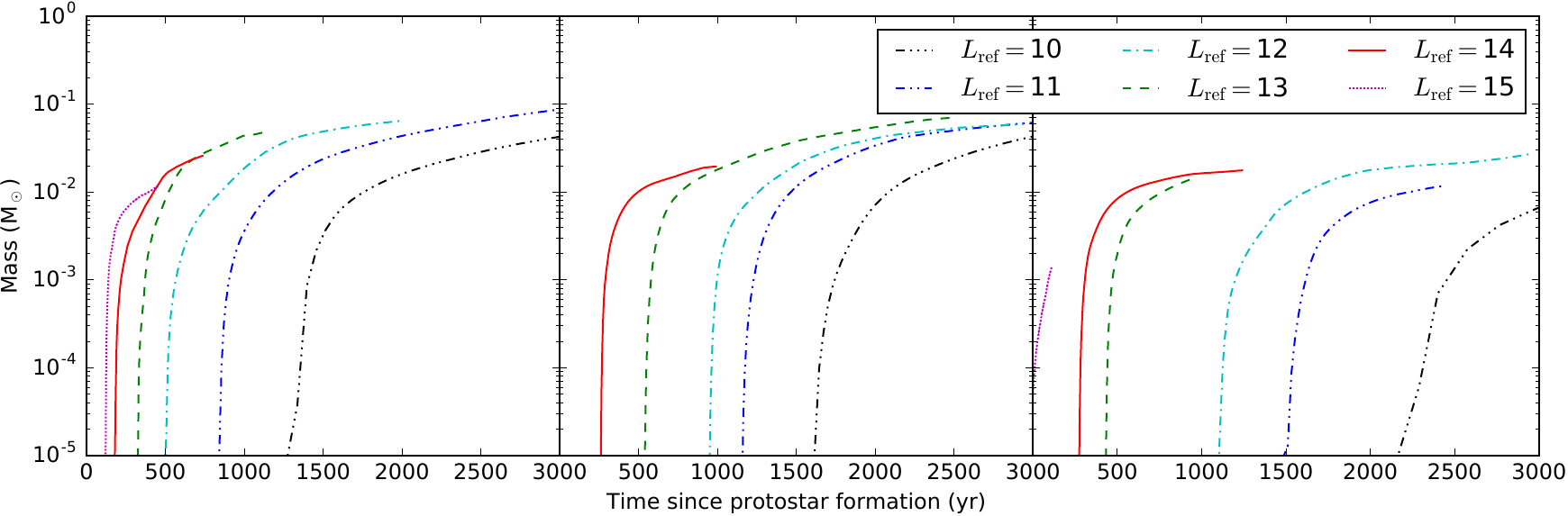}}
	\caption{Outflow mass measured as described in \Cref{ssec:outflows} for \emph{Single Star} (left), \emph{Tight Binary} (middle) and \emph{Wide Binary} (right). This was measured for levels of refinement of $L_\mathrm{ref} = 10$ (dash-dash-dash-dot line), $L_\mathrm{ref} = 11$ (dash-dash-dot line), $L_\mathrm{ref} = 12$ (dash-dot line), $L_\mathrm{ref} = 13$ (dashed line), $L_\mathrm{ref} = 14$ (solid line) and $L_\mathrm{ref} = 15$ (dotted line)}
	\label{outflow_mass}
\end{figure*}

\begin{figure*}
\centerline{\includegraphics[width=1.0\linewidth]{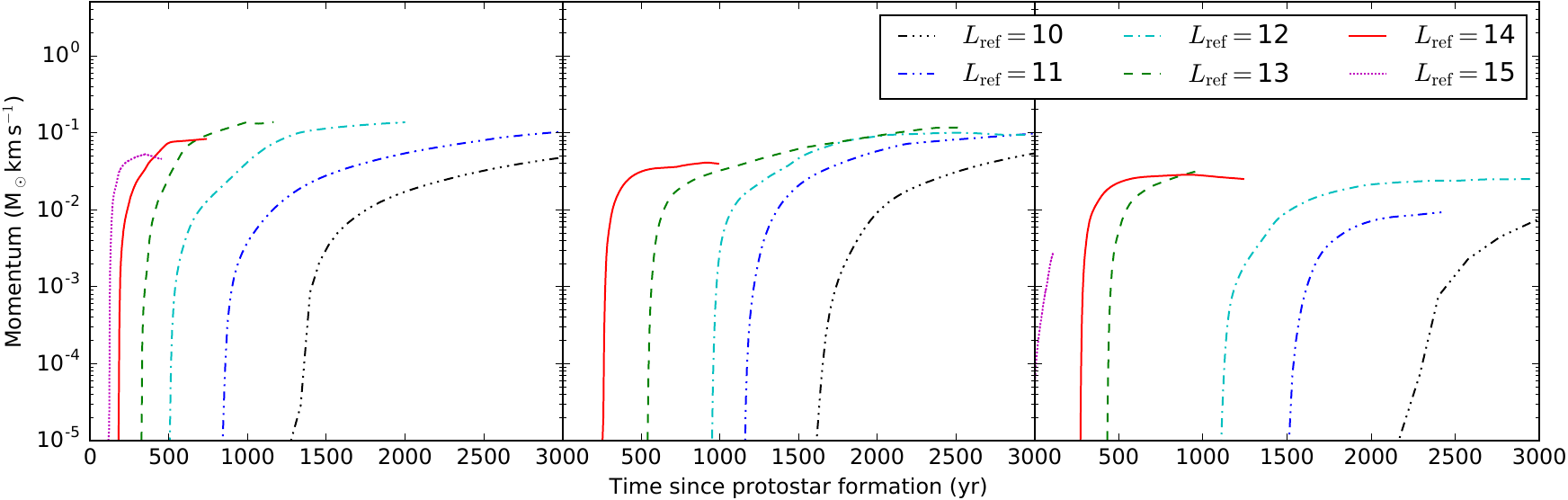}}
	\caption{Same as \Cref{outflow_mass} but for the linear momentum carried by the outflow mass}
	\label{outflow_momentum}
\end{figure*}

\begin{figure*}
\centerline{\includegraphics[width=1.0\linewidth]{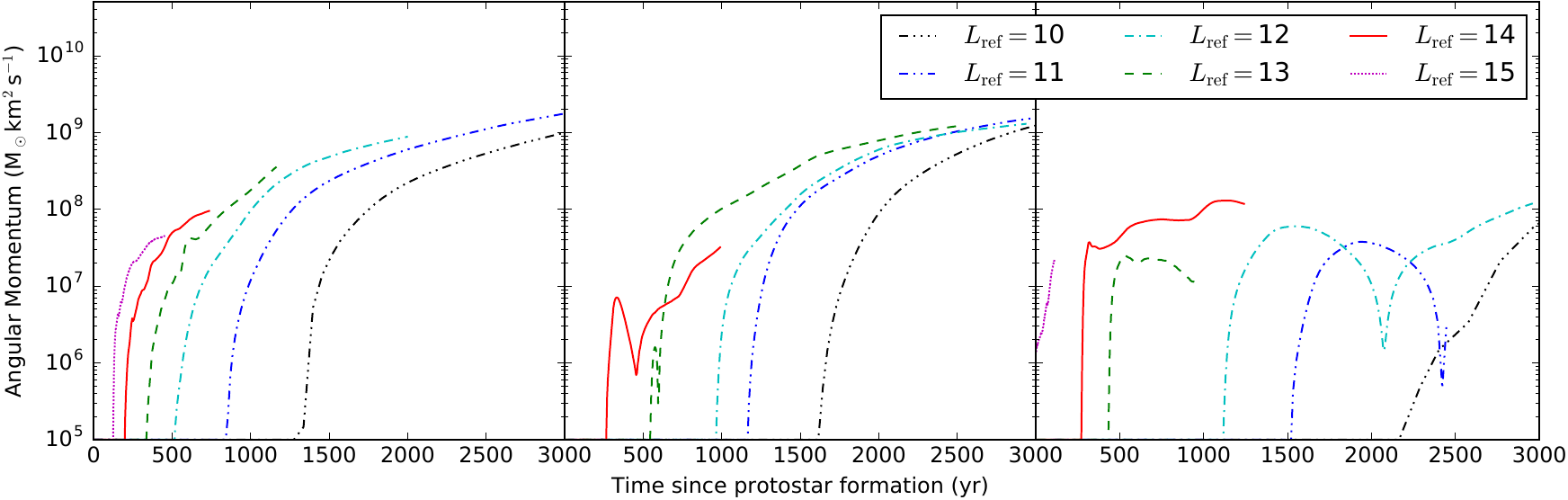}}
	\caption{Same as \Cref{outflow_mass} but for the angular momentum carried by the outflow mass}
	\label{outflow_ang_momentum}
\end{figure*}

\begin{figure*}
\centerline{\includegraphics[width=1.0\linewidth]{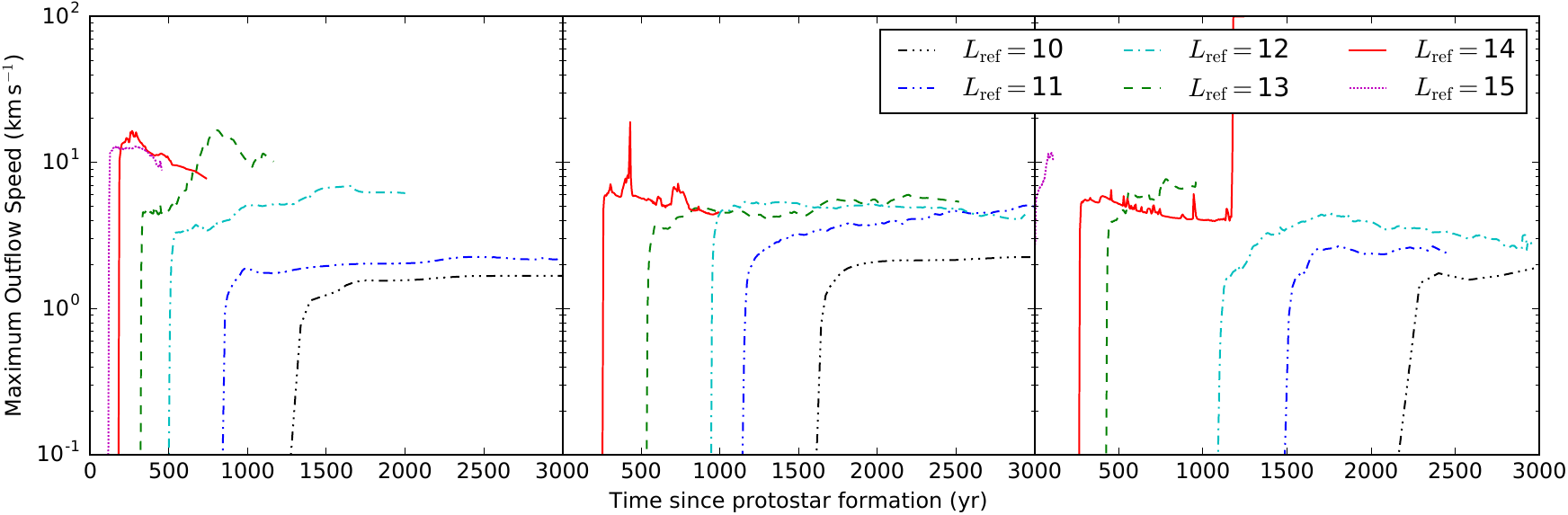}}
	\caption{Same as \Cref{outflow_mass} but for the maximum velocity of the outflow mass.}
	\label{outflow_maximum_speed}
\end{figure*}

Here we present the resolution study carried out prior to our research of the three scenarios in order to determine an appropriate resolution to carry out the study. Here we measured the outflow mass (\Cref{outflow_mass}), momentum (\Cref{outflow_momentum}) and angular momentum (\Cref{outflow_ang_momentum}) of outflow mass and maximum speed (\Cref{outflow_maximum_speed}) of outflow mass by measuring the quantities flowing through two measuring cylinders. The cylinders have radii and height of $500\au$, and are placed $250\au$ above and below the disc midplane. Within the volumes outflow mass is defined as any mass in cells with $v_z>0$ for $z>0$ and $v_z<0$ for $z<0$. From the outflow mass, the angular momentum and linear momentum of the outflows is calculated. The linear momentum is calculated from the magnitude of the velocity and the outflow mass. The angular momentum is calculated about the centre of mass of the systems. . 

When looking at these resoluation studies we see that the simulations have not converged. This is not a major issue as the overall purpose of this study is to determine the influence binarity may have on outflow morphology and momentum transport efficiencies.

When we look at \Cref{outflow_maximum_speed} we see that the maximum outflow speed increases and the outflow mass reaches the measuring volume earlier with greater resolution. This is expected, because with higher resolution we are resolving the inner regions of the disc where higher velocity winds are launched. We see this trend of increasing velocity in all three cases.

When looking at the other measured quantities in this resolution study, we see that lower resolution simulation ($L_\mathrm{ref} =10,11$) do not appear to reach a saturation level in the $3000\yr$ after protostar formation. However the simulations with resolution $L_\mathrm{ref} \gtrsim 12$, despite not reaching a plateau in their runtimes, have converged towards the measured quantites of the previous level of refinment in the duration of the simulation.

A level of refinement of $L_\mathrm{ref} = 12$ is chosen in our simulations as it allows us to run a simulation for a few thousand years after protostar formation which also resolves a strong jet component in \emph{Single Star}.

With \emph{Wide Binary}, when looking at the angular momentum we see a dip which is occurs when the sink particles are at periastron after sink particle formation. There is some resolution dependence on this dip because with higher resolution the sink particles form sooner and fall into each other, towards periastron at earlier times.

\end{document}